\input harvmac
\input psfig
\newcount\figno

\figno=0
\def\fig#1#2#3{
\par\begingroup\parindent=0pt\leftskip=1cm\rightskip=1cm\parindent=0pt
\global\advance\figno by 1
\midinsert
\epsfxsize=#3
\centerline{\epsfbox{#2}}
\vskip 12pt
{\bf Fig. \the\figno:} #1\par
\endinsert\endgroup\par
}
\def\figlabel#1{\xdef#1{\the\figno}}
\def\encadremath#1{\vbox{\hrule\hbox{\vrule\kern8pt\vbox{\kern8pt
\hbox{$\displaystyle #1$}\kern8pt}
\kern8pt\vrule}\hrule}}
\def\underarrow#1{\vbox{\ialign{##\crcr$\hfil\displaystyle
 {#1}\hfil$\crcr\noalign{\kern1pt\nointerlineskip}$\longrightarrow$\crcr}}}
%
\overfullrule=0pt

%
\def\tilde{\widetilde}
\def\bar{\overline}
\def\Z{{\bf Z}}
\def\T{{\bf T}}
\def\S{{\bf S}}
\def\R{{\bf R}}

\font\zfont = cmss10 

\def\bigone{\hbox{1\kern -.23em {\rm l}}}
\def\ZZ{\hbox{\zfont Z\kern-.3emZ}}
\def\RR{\hbox{\zfont R\kern-.3emR}}\def\TT{\hbox{\zfont T\kern-.1emT}}
\def\SS{\hbox{\zfont S\kern-.3emS}}

\Title{hep-th/0201018} {\vbox{\centerline{Deconstruction, $G_2$
Holonomy, }
\bigskip
\centerline{and Doublet-Triplet Splitting }}}
\smallskip
\centerline{Edward Witten}
\smallskip
\centerline{\it Institute For Advanced Study, Princeton NJ 08540 USA}


\medskip

\noindent
We describe a mechanism for using discrete symmetries to solve
the doublet-triplet splitting problem of four dimensional
supersymmetric GUT's.  We present two versions of the mechanism,
one via ``deconstruction,'' and one in terms of $M$-theory
compactification to four dimensions on a manifold of $G_2$
holonomy. \Date{October, 2001}

\newsec{Introduction}

One of the central problems in four-dimensional Grand Unified
Theories (GUT's) is the splitting between standard model Higgs
doublets and their color triplet partners.  The problem persists
in supersymmetric GUT's, which will be the focus of the present
paper.

In $SU(5)$ models, for example, the Higgs doublets can be
naturally placed in chiral superfields $V,\,\tilde V$ transforming
as ${\bf 5}\oplus \bar{\bf 5}$.  $V$ has for its standard model
content a possible Higgs doublet $H$ as well as a color triplet
$Q$, while $\tilde V$ has fields $\tilde H$, $\tilde  Q$
transforming in the conjugate representations.  To get standard
model phenomenology, $H$ and $\tilde H$ must be essentially
massless at the GUT scale -- receiving mass only at the
electroweak scale. But $Q$ and $\tilde Q$ have renormalizable
couplings -- related by $SU(5)$ to the couplings of $H$ and
$\tilde H$ that are needed to give mass to quarks and leptons --
that mediate baryon number violating processes. $Q$ and $\tilde
Q$ must therefore obtain masses close to the GUT scale in order
to obtain an even roughly reasonable proton lifetime.  (Even if
this is achieved, there are more obstacles to getting a realistic
proton lifetime; they will be discussed in section 2.1.)

A variety of field theory solutions to the doublet-triplet
splitting problem or fine-tuning problem have been proposed.  For
a brief review of some of the proposals up to 1995, see
\ref\randall{L. Randall and Csaba Cs\'aki, ``The Doublet-Triplet
Splitting Problem And Higgses As Pseudo-Goldstone Bosons,''
hep-ph/9508208, in {\it Particles, Strings, and Cosmology}, ed.
J. Bagger (World-Scientific, 1996).}. There also are more recent
field theoretic proposals such as one based on strong
supersymmetric dynamics \ref\kitano{R. Kitano and N. Okada,
``Dynamical Doublet-Triplet Higgs Mass Splitting,''
hep-ph/0105220.}.

Possible solutions to the problem also exist in the framework of
string theory and higher dimensions \ref\chsw{P. Candelas, G.
Horowitz, A. Strominger, and E. Witten, ``Vacuum Configurations
For Superstrings,'' Nucl. Phys. {\bf B258} (1985) 46.}. In this
context, unification only arises in some dimension greater than
four and the unified group $G$ is broken down to the standard
model (or an extension of the standard model that is
phenomenologically viable at relatively low energies) in the
process of compactification. A key ingredient in this approach is
``gauge symmetry breaking by Wilson lines,'' in which one aims,
while compactifying from ten to four dimensions, to project the
dangerous color triplets out of the low energy spectrum while
leaving the Higgs doublets.\foot{A generalization of symmetry
breaking by Wilson lines is symmetry breaking by orbifolds
\ref\dhvw{L. Dixon, J. A. Harvey, C. Vafa, and E. Witten,
``Strings on Orbifolds,'' Nucl. Phys. {\bf B261} (1985) 678.},
where the symmetry breaking is carried out using a discrete
symmetry that does not act freely. In perturbative string theory,
this is not usually used as a mechanism for GUT symmetry
breaking, because typically the massless twisted sector modes
would not be in complete $G$ multiplets and the successes of
grand unification would not be preserved.}   For a review of
Calabi-Yau compactification of the heterotic string, in which
this mechanism can naturally be incorporated, see chapters 14-16
of \ref\gsw{M. B. Green, J. H. Schwarz, and E. Witten, {\it
Superstring Theory}, Vol. 2 (Cambridge University Press, 1987).}.
For a more recent discussion of some stringy constructions, see
\ref\faraggi{A. Faraggi, ``Doublet Triplet Splitting In Realistic
Heterotic String Derived Models,'' hep-ph/0107094, Phys. Lett.
{\bf B520} (2001) 337.}.  Mechanisms of roughly this type have
lately come to be widely studied from a more phenomenological and
bottom-up
 \nref\ak{Y. Kawamura, ``Triplet Doublet Splitting,
Proton Stability and Extra
Dimension,'' Prog. Theor. Phys. {\bf 105} (2001) 999, hep-ph/0012125.}%
\nref\bk{L. J. Hall and Y. Nomura, ``Gauge Unification In Higher
Dimensions,'' Phys. Rev. {\bf D64} (2001) 055003, hep-ph/010125.}%
\nref\eek{G. Altarelli and F. Feruglio, ``$SU(5)$ Grand
Unification In Extra Dimensions And Proton Decay,'' Phys. Lett.
{\bf B511} (2001) 257.}%
\nref\egek{M. Kakizaki and M. Yamaguchi, ``Splitting Triplet And
Doublet In Extra Dimensions,'' hep-ph/0104103.}%
 \nref\ck{L. J. Hall, H. Murayama, and Y. Nomura,
``Wilson Lines
And Symmetry Breaking On Orbifolds,'' hep-th/0107245.}%
 point of view \refs{\ak -\ck}.

\def\5{{\bf 5}}
In the present paper, we will be concerned with solutions to the
fine-tuning problem that make use of discrete symmetries.  In
fact, some but not all field theory proposals for the fine-tuning
problem and some but not all string theory proposals make use of
discrete symmetries.\foot{Discrete symmetries were not used, for
example, in the doublet-triplet splitting mechanism proposed in
\chsw.} The basic reason that discrete symmetries might be
relevant to the fine-tuning problem, in a supersymmetric GUT-like
theory, is as follows.

Suppose that we are given a discrete symmetry $F$ under which the
components $(Q,H)$ of the $\5$ transform as
$(e^{i\alpha},e^{i\beta})$, while the components $(\tilde Q,\tilde
H)$ of the $\bar \5$ transform as $(e^{i\gamma},e^{i\delta})$. If
$e^{i(\alpha+\gamma)}=1$, then this symmetry allows $Q$ and
$\tilde Q$ to get GUT scale masses, while if
$e^{i(\beta+\delta)}\not= 1$, then $H$ and $\tilde H$ are
massless. In fact, in this scenario, the ``$\mu$-term,'' an
$H\tilde H$ term in the superpotential that is needed for
supersymmetric phenomenology, violates $F$ and can only arise at
lower energies where (hopefully) $F$ is spontaneously broken.

{}From a field theory point of view, it can be difficult,
depending on one's assumptions, to get a discrete symmetry with
the necessary properties. It is generally not true in GUT's that a
discrete symmetry of the low energy theory must commute with the
GUT group $G$; it might be the product of a discrete symmetry that
``normalizes'' the standard model subgroup of $G$ (conjugates it
to itself) times an ordinary discrete symmetry that commutes with
$G$. However, if $G=SU(5)$, an element of $G$ that normalizes
$SU(3)\times SU(2)\times U(1)$ is actually contained in
$SU(3)\times SU(2)\times U(1)$.  This means that, modulo a
standard model gauge transformation, a discrete symmetry in a
four-dimensional $SU(5)$ model actually commutes with $SU(5)$.  A
discrete symmetry that is the product of a standard model gauge
transformation and a symmetry that commutes with $SU(5)$  leaves
the $H\tilde H$ term in the superpotential  invariant if and only
if the $Q\tilde Q$ term is invariant.  (Both terms are invariant
under the standard model gauge group, and a discrete symmetry
that commutes with $SU(5)$ does not distinguish them either.)  So
such a discrete symmetry cannot solve the fine-tuning problem.

Things are no different in four-dimensional GUT's based on the
other standard simple GUT groups such as $SO(10)$ and $E_6$.  The
reason is that each of these groups, with the usual standard
model embedding, contains a unique $SU(5)$ subgroup $G'$ that
contains $SU(3)\times SU(2)\times U(1)$, and the above argument
can be carried out using $G'$.

As explained in \ref\gw{M. Goodman and E. Witten, ``Global
Symmetries In Four Dimensions And Higher Dimensions,'' Nucl.
Phys. {\bf B271} (1986) 21.}, in the context of a four-dimensional
$SU(5)$ model, mixing the $\5$ and $\bar\5$ with additional
$SU(5)$ representations does not change this conclusion, but if
one starts above four dimensions, one can readily get discrete
symmetries of the desired kind.

This is a benefit of having extra dimensions.  However, it has
recently been pointed out \nref\georgietal{N. Arkani-Hamed, A. G.
Cohen, and H. Georgi, ``(De)constructing Dimensions,''
hep-th/0104005, Phys. Rev. Lett. {\bf 86} (2001) 4757. }%
\nref\hill{C. T. Hill, S. Pokorski, and J. Wang, ``Gauge Invariant
Effective Lagrangian For Kaluza-Klein Modes,'' hep-th/0104035.}%
\refs{\georgietal,\hill}
 that some
higher-dimensional setups can be ``deconstructed,'' or simulated
by a four-dimensional model in which, roughly speaking, the extra
dimensions are replaced by a lattice (which may have a very small
number of lattice points). In section 2, following this lead, we
deconstruct one version of the higher-dimensional approach to the
doublet-triplet splitting problem.  In its minimal form, this
involves beginning with an $SU(5)\times SU(5)$ gauge theory, with
the standard model diagonally embedded in the product of the two
$SU(5)$'s.  Such structures have been considered in many papers
on deconstruction such as \nref\forex{N. Arkani-Hamed, A. G.
Cohen, and H. Georgi,
``Accelerated Unification,'' hep-th/0108089.}%
\refs{\georgietal,\forex}.     By starting with $SU(5)\times
SU(5)$ rather than $SU(5)$, the constraints on discrete
symmetries are relaxed, and it is readily possible to find
discrete symmetries that can solve the fine-tuning problem.

In section 3, we present a higher-dimensional version of the same
mechanism.  In fact, we describe how discrete symmetries that can
naturally split triplets from doublets can arise in the context
of $M$-theory compactification to four dimensions on a manifold
of $G_2$ holonomy.  This is a natural way to obtain a
four-dimensional model with ${\cal N}=1$ supersymmetry, and since
it can be dual to heterotic string compactification on a
Calabi-Yau threefold, it is fairly clear that it must be possible
to express some of the mechanisms for doublet-triplet splitting
that are familiar for the heterotic string in the language of
compactification on $G_2$ manifolds.  We do this in section 3.
Some of the ingredients of this construction have appeared in
previous papers
\nref\ewitten{E. Witten, ``Anomaly
Cancellation on $G_2$ Manifolds,'' hep-th/0108165.}%
\nref\baw{B. Acharya and E. Witten, ``Chiral Fermions From Manifolds Of
$G_2$ Holonomy,'' hep-th/0109152.}%
\refs{\ewitten,\baw}.

In fact, the approach sketched in section 3 was worked out first.
Deconstruction was attempted following a question raised by
Hsin-Chia Cheng, when this work was presented in a seminar at the
University of Chicago.

\bigskip\noindent
{\it Similarities And Differences Of The Two Approaches}

The deconstructed theory presented in section 2 is not technically
a unified theory by some definitions, since the $SU(5)\times
SU(5)$ gauge theory has two independent gauge couplings (there is
no symmetry exchanging the two factors).  However, the diagonal
embedding of the standard model ensures that most of the familiar
consequences of grand unification, such as the SUSY-GUT
prediction for $\sin^2\,\theta_W$  and constraints on the quantum
numbers of quarks and leptons, do hold. By contrast, the
$M$-theory model unifies not just the gauge fields but also the
Higgs fields, standard model fermions, and gravity.

The two types of model differ in a few other interesting ways. In
$M$-theory models, it is believed (there is not a complete proof
of this) that the discrete symmetries are always anomaly-free
(they may be spontaneously broken by the transformation law of an
axion). In the context of deconstruction, opinions may differ
about whether an anomalous discrete symmetry should be considered
technically natural, but at any rate it would be
phenomenologically viable to try to solve the fine-tuning problem
using such a  symmetry.

In the deconstructed model, gauge anomalies must cancel
separately in each $SU(5)$ factor of the gauge group.  Gauge
anomaly cancellation is a less severe constraint in the
$M$-theory approach, since there is an anomaly inflow mechanism
\ewitten\ (analogous to anomaly inflow for $D$-branes
\ref\mbetal{M. B. Green, J. A. Harvey, and G. W. Moore,
``$I$-Brane Inflow And Anomalous Couplings On $D$-Branes,''
hep-th/9605033, Class. Quant. Grav. {\bf 14} (1997) 47.}) that
can shift the anomaly from one factor to the other. Anomaly
inflow, since it involves Chern-Simons-like couplings, appears
difficult to deconstruct, but see \ref\skiba{W. Skiba and D. Smith,
``Localized Fermions And Anomaly Inflow Via Deconstruction,'' hep-ph/0201056}
(which appeared on hep-ph a few days after the original version of the
present paper).   Finally, like most perturbative
heterotic string models \ref\ww{X.-G. Wen and E. Witten,
``Electric And Magnetic Charges In Superstring Models,'' Nucl.
Phys. {\bf B261} (1985) 651.}, the models derived from $M$-theory
generally have superheavy unconfined color singlet particles with
fractional electric charges, and reciprocally a larger quantum of
magnetic charge than would be expected in a four-dimensional
GUT.  The deconstructed models obey conventional quantization of
electric and magnetic charge.

After submission of the original version of the present paper, I learned
of \ref\barbieri{R. Barbieri, G. Dvali, and A. Strumia,
``Strings Versus Supersymmetric GUT's: Can They Be Reconciled?'' Phys.
Lett. {\bf B333} (1994) 79,
hep-ph/9404278.}, which presents a construction similar to that in section 
2.1.

\newsec{$SU(5)'\times SU(5)''$ And Deconstruction}

\subsec{Direct Construction Of The Model}

We start with a gauge theory in four-dimensions in which the
gauge group is the product $G=SU(5)'\times SU(5)''$ of two copies
of $SU(5)$.  We suppose that the standard model group is
diagonally embedded in the product of the two factors. The
hypercharge group $U(1)_Y$, for example, is the diagonal subgroup
of $U(1)_Y'\times U(1)_Y''$, the product of the hypercharge groups
of the two $SU(5)$'s.

We assume that, in addition to the standard model being unbroken,
a discrete global symmetry group  $F'\cong \Z_n$ is unbroken at
the GUT scale.  We take $F'$ to be a diagonal product of an
ordinary global symmetry $F=\Z_n$ (which commutes with $G$) and
the $\Z_n$ subgroup of $U(1)_Y''$. In what follows, we pick a
fixed generator of $F'$. An explicit and fairly simple set of
Higgs fields that can break $G\times F$ to $SU(3)\times
SU(2)\times U(1)\times F'$ will be given in section 2.2.

\def\w1{{\bf 1}}
Given this low energy structure, it is straighforward to solve
the doublet-triplet splitting problem.  We suppose that the Higgs
bosons, whose expectation values will ultimately give masses to
quarks and leptons, consist of multiplets $V,\, \tilde V$
transforming under $SU(5)'\times SU(5)''$ as
$(\5,\w1)\oplus(\w1,\bar\5)$. $V$ decomposes under the standard
model as $(Q,H)$ and $\tilde V$ decomposes as $(\tilde Q,\tilde
H)$; here (as in the introduction) $H$ and $\tilde H$ are standard
model Higgs fields and $Q,\,\tilde Q$ are colored partners.

$V$ is neutral under $U(1)_Y''$, so in this multiplet $F'$ acts as
an ordinary global symmetry.  Hence, under the generator of $F'$,
$V$ transforms as
 \eqn\exfac{\left(\matrix{ Q\cr H\cr}\right)\to e^{i\alpha}
 \left(\matrix{ Q\cr H\cr}\right),}
 for some $\alpha$.  But on the $(\w1,\bar \5)$, $F'$ acts as the
 product of a global symmetry and a $U(1)_Y''$ gauge
 transformation, so the transformation under the generator of $F'$ is
  \eqn\xfac{\left(\matrix{ \tilde Q\cr \tilde H\cr}\right)\to
 \left(\matrix{ e^{i\gamma}\tilde Q\cr e^{i\delta}\tilde
 H\cr}\right).} Here $e^{i\gamma}$ and $e^{i\delta}$ are
 arbitrary $n^{th}$ roots of 1, depending on the choice of $F$
 charge of the $(\w1,\bar\5)$ as well as the precise diagonal
 subgroup of $F\times U(1)_Y''$ we have chosen for $F'$.  Now it
 is clear how to solve the doublet-triplet splitting problem; we
 merely choose the charges so that $e^{i(\alpha+\gamma)}=1$ but
 $e^{i(\alpha+\delta)} \not= 1$.  Then a $Q\tilde Q$ term in the
 superpotential is $F'$-invariant, but $F'$ forbids an $H\tilde H$
 term.

 Now let us consider how to incorporate quarks and leptons in this
 model.  There are many choices, as the standard model quantum
 numbers of quarks and leptons could originate from either or both
 of the two $SU(5)$'s.  We consider two illustrative models:

\def\10{{\bf 10}}

\bigskip\noindent{\it All Quarks And Leptons From The First
Factor}

  In our first model, we assume that all quark and lepton
 quantum numbers arise from the first $SU(5)$.  So the quarks and
 leptons arise from three copies of $(\10,\w1)\oplus (\bar \5,\w1)$.
 $F'$ acts by ordinary $G$-invariant global symmetries on these multiplets.
 We assume that the $(\10,\w1)$'s all transform by multiplication
 by $e^{i\sigma}$, with a common $\sigma$, and likewise the $(\bar
 \5,\w1)$'s all transform by multiplication by $e^{i\tau}$ for
 some $\tau$.

 Let us see what constraints are required by phenomenology.  To
 give masses to up quarks, we want  $H (\10,\w1)^2$ superpotential
 couplings; for this, we need
 \eqn\ugg{e^{i(\alpha+2\sigma)}=1.}
 To give masses to down quarks and charged leptons, we want
 $\tilde H(\10,\w1)(\w1,\bar \5)$ interactions; for this, we need
 \eqn\nugg{e^{i(\delta +\sigma+\tau)}=1.}
 The experimental observation of neutrino masses strongly suggests
 that an $H^2(\bar\5,\w1)^2$ coupling is allowed; in the context of
 GUT's, this leads to neutrino masses of roughly the observed
 magnitude.  For this coupling to be allowed, we need
 \eqn\tugg{e^{2i(\alpha+\tau)}=1.}
 But we do not want to allow a $H(\bar \5,\w1)$ mass term, so we want
 \eqn\uggl{e^{i(\alpha+\tau)}=-1.}
 We can solve these equations in terms of an arbitrary angle
 $\sigma$:\foot{All equations for these angles are of
 course understood mod $2\pi$.}
 \eqn\juggle{\eqalign{ \alpha & = -2\sigma \cr
                       \tau   & = 2\sigma+\pi \cr
                       \delta & = -3\sigma +\pi.\cr}}
 Finally, and of great importance, to get a realistic proton
 lifetime, we need additional restrictions.  We want to avoid
 renormalizable couplings $(\10,\w1)(\bar \5,\w1)^2$ that violate baryon
 number, so we want $e^{i(\sigma+2\tau)}\not= 1$, or in terms of
 the above solution
 \eqn\duggle{5\sigma\not= 0.}
Moreover, it is highly desireable to avoid $(\10,\w1)^3(\bar
\5,\w1)$
 terms in the superpotential, which lead to dimension five baryon
 nonconserving operators.   It is difficult for a GUT-like model
 to generate such terms and have a sufficiently long-lived proton;
 for a recent account, see \ref\mura{H. Murayama and A. Pierce, ``Not Even Decoupling
 Can Save Minimal Supersymmetric $SU(5)$,'' hep=ph/0108104. }.
 So we want $e^{i(3\sigma+\tau)}\not= 1$, or in terms of the above
 solution,
 \eqn\guggle{5\sigma+\pi\not= 0.}

 One interesting feature of this model is that although the
 $H(\10,\w1)^2$ Yukawa couplings that give masses to up quarks can
 arise from  ordinary $G$-invariant cubic terms $V(\10,\w1)^2$ in the
 superpotential, the $\tilde
 H(\10,\w1)(\bar\5,\w1)$ couplings that give the down quarks and
 charged lepton masses cannot  so
 arise, given that $\tilde H$ transforms as $(\w1,\bar \5)$.
 These latter couplings must instead be induced
 from  unrenormalizable superpotential couplings
 of the general form $\tilde V(\10,\w1)(\bar \5,\w1) \Phi$, where $\Phi$ is
 constructed from fields (introduced in section 2.2) whose
 expectation values break $G\times F$ to the low energy subgroup $SU(3)\times
 SU(2)\times U(1)\times F'$, which  does allow
 the $\tilde H(\10,\w1)(\bar \5,\w1)$ couplings.
Because they arise from unrenormalizable couplings, it is natural
to expect the down quark and charged lepton masses to be less than
the up quark masses.  For the third generation, this is true, as
the bottom quark and tau lepton are much lighter than the top
quark.  If the ratios $m_b/m_t$ and $m_{\tau}/m_t$ are really to
be obtained this way, the cutoff scale characterizing
unrenormalizable interactions cannot be too much bigger than the
GUT scale, since after all $m_b/m_t$ and $m_{\tau}/m_t$ are only
moderately small. Moreover, the fact that the first two
generations are so light compared to the top quark would
presumably require some further mechanism.

Finally, the spectrum of the model as we have presented it so far
cannot be the whole story, since it is anomalous.  The only
chiral multiplet so far introduced that carries  $SU(5)''$
charges is the Higgs multiplet $\tilde V$, transforming as
$(\w1,\bar\5)$.  The couplings of this field are anomalous.
Likewise, $SU(5)'$ couples to three anomaly-free copies of
$(\10,\w1)\oplus (\bar \5,\w1)$ as well as a Higgs multiplet $V$
transforming as $(\5,\w1)$; its couplings are again anomalous. A
simple way to cancel the anomalies is to add additional fields
$\tilde S,S$ transforming as $(\bar\5,\w1)\oplus (\w1,\5)$; their
$F$ quantum numbers should be restricted to avoid various
undesireable couplings.  If one believes that $F$ should be
anomaly-free for naturalness of the model, then the $F$ quantum
numbers of $\tilde S$, $S$ must be further constrained.  Purely
for phenomenological purposes, however, gauge anomalies in $F$
would not lead to trouble.

It is interesting to speculate that the fields $\tilde S,S$ might
play the role of ``messenger fields'' in gauge-mediated
supersymmetry breaking  (for surveys see
\nref\gmsb{M. Dine, Y.
Nir, and Y. Shirman, ``Variations on Minimal Gauge-Mediated
Supersymmetry Breaking,'' hep-ph/9607397, Phys. Rev. {\bf D55}
(1997) 1501.}%
\nref\weinberg{S. Weinberg, {\it The Quantum Theory of Fields,
Vol. 3: Supersymmetry}, (Camb. Univ. Press, 2000).}%
\refs{\gmsb,\weinberg}), communicating to the standard model
fields the occurrence of supersymmetry breaking in a hidden
sector. For this, there might be singlets $T$ whose expectation
values violate supersymmetry and $F'$ and which have
superpotential couplings $T S\tilde S$. Actually, since the color
singlet and color triplet components of $S$ transform differently
under $F'$ (and there is no such splitting for $\tilde S$), one
would need different $T$ fields transforming differently under
$F'$ to couple to the color singlets and triplets in $S\tilde S$.

\bigskip\noindent
{\it Mixed Origin Of Quarks And Leptons}

For a second model, which we will not develop as fully, we
consider one possibility among many to use the distinct gauge
theoretic origin of the two different Higgs fields to constrain
quark and lepton masses.  Since the top quark is much heavier
than the charm or up quark, we might assume that the top
originates from a $(\10,\w1)$ while the charm and up quarks
originate from two copies of $(\w1,\10)$.  Then the top quark
gets a mass from renormalizable $V(\10,\w1)^2$ couplings, while
the charm and up masses originate from unrenormalizable
interactions. Since the bottom quark and tau lepton are much
heavier than the analogous particles in the first two
generations, one might similarly suppose that the bottom quark
arises from a $(\w1,\bar \5)$ (so that it can get its bare mass
from a renormalizable coupling $\tilde V(\w1,\10)(\w1,\bar \5)$)
and the others from two copies of $(\bar \5,\w1)$. This spectrum,
including the Higgs fields, is fortuitously anomaly-free so we do
not need additional fields analogous to $S,\tilde S$ of the first
model.

A problem with this model is that it will be hard to generate
masses for all down quarks and charged leptons.  In fact, one
down quark mass and one charged lepton mass would have to come
from a coupling $\tilde H(\bar \5,\w1)(\w1,\10)$.  Because
different components of the $(\w1,\10)$ transform differently
under $F'$, while there is no such splitting for the $(\bar
\5,\w1)$, this coupling cannot give a mass to both a down quark
and a charged lepton, no matter what $F'$ charges we assume.

 \subsec{Interpretation Via Deconstruction}

Next we will explain in what sense the above model can arise via
deconstruction. First, let us explain what manifold is being
deconstructed.  We let $D_0$ be a two-dimensional disc.  We can
triangulate it as in the figure, with one vertex $P$ in the
center and $n$ vertices $Q_1,\dots,Q_n$ on the boundary.

The space we want to deconstruct is not $D_0$, but rather a space
$D$ obtained by imposing the following equivalence relation: two
points in $D_0$ that are {\it on the boundary} are equivalent if
they differ by a $2\pi/n$ rotation of the boundary.  Thus, an
equivalence relation is imposed only on the boundary.  If $n=2$,
$D$ is an unorientable manifold, the real projective plane ${\bf
RP}^2$.  Its deconstruction was described in \ref\otheracg{N.
Arkani-Hamed, A. Cohen, and H. Georgi, ``Twisted Supersymmetry
And The Topology Of Theory Space,'' hep-th/0109082.} in the
discussion of ``spider web theory space.''  The case $n=2$ would
not quite work for us, at least in the first model presented
above, because \guggle\ and \duggle\ could not be obeyed.  The
deconstruction, however, is similar for $n>2$,  though $D$ is not
a manifold (but a singular topological space) for $n>2$. In fact,
the triangulation or deconstruction of $D$ is very simple. There
are only two vertices, $P$ and $Q_1$ (since the equivalence
relation identifies all the $Q_i$), connected by the edges shown
in the figure.

\centerline{\psfig{figure=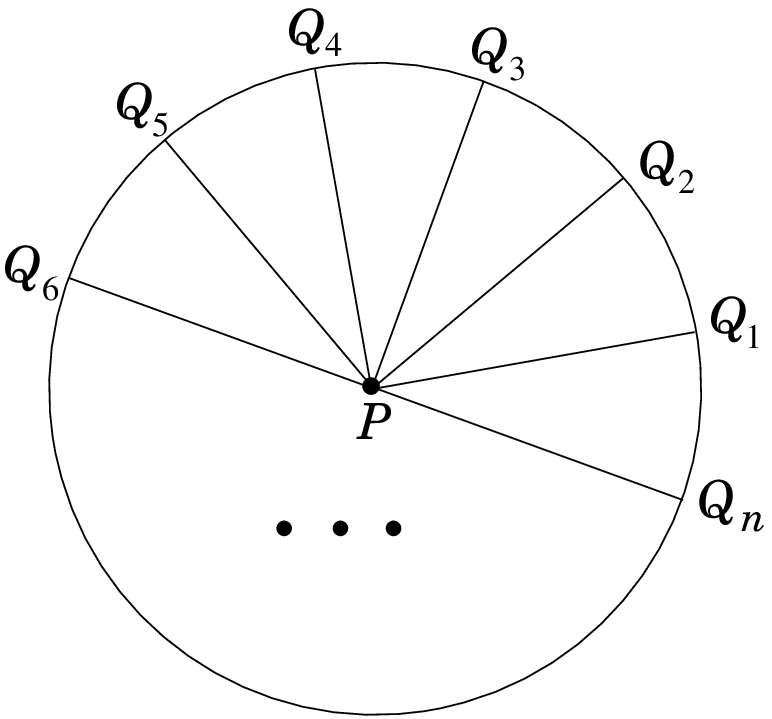,width=3in}}

\vskip 1cm
\centerline {{\it Figure 1:} Moose or quiver diagram representing
the triangulation}
 \centerline{or deconstruction of a two-dimensional disc
$D_0$.}

\vskip .5cm

Suppose that we want to do lattice gauge theory of gauge group
$G$ on $D$, understood in terms of this triangulation. The
lattice gauge field that lives on the link in figure 1 from $Q_i$
to $Q_{i+1}$ is a group element $W$; because of the equivalence
relation,  $W$ is independent of $i$.  Similarly, on the link
from $P$ to $W_i$, the lattice gauge field is a group element
$U_i$. To minimize the energy, we want the holonomy around each
two-dimensional plaquette to equal 1. For the triangulation in
the figure, this means that for each $i$ we need
\eqn\kolop{U_iWU_{i+1}^{-1}=1.} Taking the product of these
relations for $i=1,\dots, n$, we deduce that $W^n=1$.  It is
possible (by making a gauge transformation at $P$) to pick
$U_1=1$, and then \kolop\ determines the other $U_i$.  So the
$G$-symmetry breaking is determined entirely by the choice of
$W$. This result really reflects the fact that a lattice gauge
field with trivial holonomy around every plaquette gives a
representation of the fundamental group, which for $D$ is
$\Z_n$.  Finiteness of the fundamental group results means that
symmetry breaking by Wilson lines  depends on discrete choices
\refs{\chsw,\otheracg}, in this case the choice up to conjugacy
of $W\in G$ with $W^n=1$.

We interpret $F\cong \Z_n$ as the discrete symmetry that rotates
the disc in figure (1) by an angle $2\pi/n$.  Note that $F$ is a
non-trivial symmetry of $D$.  $F$ might sound suspiciously similar
to the equivalence relation that was imposed on $D_0$ to get
$D$.  The difference is that the equivalence relation was imposed
only on the boundary of $D_0$, while $F$ acts on all of $D$.  So
$F$ acts non-trivially on $D$, leaving fixed the point $P$ and
also the boundary, where $F$ generates the equivalence relation.

Now, let us ask if the choice of a gauge field obeying \kolop\
leaves $F$ unbroken.  In fact, it does.  Even in the presence of
the gauge fields, $F$ is unbroken if accompanied by the gauge
transformation $W$ at the site $Q_1$ (and no gauge transformation
at $P$).  Under such a gauge transformation, $W$ is invariant and
$U_i$ transforms to $U_iW$, which by \kolop\ equals $U_{i+1}$.
That is what $U_i$ should transform into under the $2\pi/n$
rotation. The combined global symmetry $F$ plus gauge
transformation $W$ will be called $F'$.

 To make contact with section 2.1, suppose
$G=SU(5)$.  Then in four-dimensional terms, the gauge group is
$SU(5)'\times SU(5)''$, where one factor ``lives'' at $P$ and the
other at $Q_1$.  The Wilson lines described above break
$SU(5)'\times SU(5)''$ to the subgroup of $SU(5)$ that commutes
with $W$. Of course, if \eqn\umptly{W=\left(\matrix{e^{2i\rho} &
& & & \cr
                                        & e^{2i\rho} & & & \cr
                                        & & e^{2i\rho} & & \cr
                                        & & & e^{-3i\rho} & \cr
                                         & & & & e^{-3i\rho}\cr
                                         }\right),}
for some suitable $n^{th}$ root of unity $e^{i\rho}$, then the
unbroken group is the standard model subgroup $SU(3)\times
SU(2)\times U(1)$.

The lattice gauge theory construction just sketched can (as in
other discussions of deconstruction) be reinterpreted in terms of
a set of Higgs fields that break $SU(5)'\times SU(5)''\times F$
down to $SU(3)\times SU(2)\times U(1)\times F'$. In fact, in moose
\ref\howd{H. Georgi, ``A Toolkit For Builders Of Composite
Models,''  Nucl. Phys. {\bf B266} (1986) 274. } or quiver
\ref\dm{M. R. Douglas and G. W. Moore, ``$D$-Branes, Quivers, And
ALE Instantons,'' hep-th/9603167. } theory, a unitary gauge group
is placed at each lattice site and a ``bifundamental'' field
$\Phi$, which in the present supersymmetric context would be a
chiral superfield, is placed on each link.  The expectation
values of the $\Phi$'s need not in general be unitary matrices,
but if the potential is suitable they will be unitary matrices
obeying \kolop\ (or they will be more general matrices leaving
the same unbroken symmetries).

To complete the construction of a model, we place additional
matter fields at the sites $P$ and $Q_1$. For example, in the
first model of section 2.1, the quarks and leptons all originate
from fields placed at $P$, while fields $S,\tilde S$ which might
be interpreted as the messenger fields of gauge-mediated
supersymmetry breaking are placed at $P$ and $Q_1$.

\bigskip\noindent{\it Analog With $D$ A Smooth Manifold}

As preparation for the next section, we will describe how one
would obtain a similar model based on deconstruction of a smooth
manifold rather than the singular space $D$.

 The only important properties
of $D$ that we used in the above was that it has fundamental group
$\Z_n$ and has a discrete symmetry $F$ of a suitable sort.
  For $n>2$, a manifold of fundamental group $\Z_n$ has at least
dimension three.  It is easy to give an example in dimension
three.  We start with a three-sphere $Q_0$, which we represent by
complex numbers $z_1,z_2$, with $|z_1|^2+|z_2|^2=1$. Then we
consider the  group $L\cong\Z_n$ that acts by
\eqn\kino{L:\,z_i\to e^{2\pi i/n}z_i,~~i=1,2.}  Since $L$ acts
freely, the quotient is a manifold $Q=Q_0/L=\S^3/\Z_n$ with
fundamental group $\Z_n$.

So in $SU(5)$ gauge theory, or lattice gauge theory, on $Q$, a
configuration of minimal energy is determined by a holonomy
matrix $W$ with $W^n=1$.  With $W$ chosen as in \umptly, $SU(5)$
is broken to $SU(3)\times SU(2)\times U(1)$.  Moreover, $Q$
admits the action of a global symmetry $F\cong \Z_n$ that acts by
\eqn\knonp{F:\,z_1\to z_1,~z_2\to e^{2\pi i/n}z_2.}

A flat gauge field on $Q$ with holonomy $W$ is $F$-invariant
modulo a gauge transformation. To describe the necessary gauge
transformation, or at least its key properties, in detail, let us
look at the fixed points of $F$. These consist of two circles
$S_1$ and $S_2$, where $S_1$ is defined by $|z_1|=1$, $z_2=0$,
and $S_2$ is defined by $z_1=0$, $|z_2|=1$. $S_1$ is obviously
left fixed by $F$, and $S_2$ is also left fixed modulo the
equivalence relation \kino.  We have defined $Q$ as $Q_0/L$,
where, to generate a flat vacuum gauge field with holonomy $W$,
when we divide by $L$ we make a gauge transformation by $W$. This
tells us how $F$ must act on the $SU(5)$ gauge bundle. Since $F$
leaves fixed $S_1$ trivially, we can take $F$ to act trivially on
the fibers of the gauge bundle over $S_1$.  But since $F$ leaves
$S_2$ fixed modulo an element of  $L$ that is accompanied by a
gauge transformation $W$, the transformation \knonp\ must be
accompanied on the fibers at $S_2$ by a gauge transformation by
$W$.  So the unbroken symmetry $F'$ is equivalent to $F$ at
$S_1$, but it is equivalent to $F$ times the gauge transformation
$W$ at $S_2$.

Now we can construct a model.  We postulate chiral superfields
that are localized at points on $Q$, as opposed to gauge fields
that propagate throughout $Q$.  To be more precise, we localize
the chiral superfields at points on $S_1$ or $S_2$, so as to
preserve the $F'$ symmetry.\foot{One could also consider including
chiral superfields localized at a set of points making a
non-trivial $F'$ orbit, but we will omit this generalization.}

To imitate the previous construction, we place Higgs fields
transforming as $\5$ of $SU(5)$ on $S_1$, and we place Higgs
fields transforming as $\bar \5$ of $SU(5)$ on $S_2$. We give the
$\5$ and $\bar\5$ any $F$ quantum numbers we want. Then under the
unbroken discrete symmetry $F'$, the doublet and triplet in the
$\5$ transform the same way under $F'$, but the doublet and
triplet in the $\bar\5$ transform differently. This leads to the
mechanism of doublet-triplet splitting that we have exploited
earlier.  By placing all quarks and leptons in $\10$'s and $\bar
\5$'s that are supported at points on $S_1$, we can imitate the
first model of section 2.1; by placing some on $S_1$ and some on
$S_2$, we can imitate the second model.

All this is in parallel with the two-dimensional example
corresponding to the picture of figure 1. The circles $S_1$ and
$S_2$ correspond respectively in that example to the point $P$
and the boundary of the disc. In fact, those were the fixed
points (modulo the equivalence relation) of the $2\pi /n$ rotation
that played the role of $F$ in the two-dimensional case.

Deconstruction of the three-dimensional model could be
accomplished by picking an $F$-invariant triangulation and using
lattice gauge theory. Instead of going in that direction, we will
now proceed to $M$-theory, where the type of construction that we
have just sketched can arise naturally in compactification on a
manifold of $G_2$ holonomy.

\newsec{Analog In $M$-Theory On A Manifold Of $G_2$ Holonomy}

\def\N{{\cal N}}
It was predicted a quarter century ago that a maximal
supergravity theory would exist in eleven dimensions \ref\nahm{W.
Nahm, ``Supersymmetries And Their Representations,'' Nucl. Phys.
{\bf B135} (1978) 149.}, and the theory was constructed
relatively soon afterwards \ref\cjs{E. Cremmer, B. Julia, and J.
Scherk, ``Supergravity Theory In Eleven Dimensions,'' Phys. Lett.
{\bf B76} (1978) 409. }. Moreover, it has been known for nearly as
long that $\N=1$ supersymmetry in four dimensions would arise in
compactifying from eleven to four dimensions on a compact
seven-manifold $X$ of $G_2$ holonomy.

This seems like an interesting starting point for making a model
of the real world, which is certainly exceptional, and where there
are hints of low energy supersymmetry. $G_2$, which is the
smallest of the compact exceptional Lie groups, of rank two and
dimension fourteen, is the only exceptional group that can be the
holonomy of a (non-homogeneous) Riemannian manifold. Moreover, in
the classification of holonomy groups of (non-homogeneous)
Riemannian manifolds, there are several infinite series, and two
exceptions -- $G_2$ in dimension seven and $Spin(7)$ in dimension
eight. Thus $G_2$ holonomy is somewhat analogous to $SU(3)$
holonomy, which is used in supersymmetric compactifications of
string theory from ten to four dimensions, but is even more
special.

All these facts suggest that as a starting point for models of
particle physics, one should seriously consider compactification
of  eleven-dimensional supergravity or its refinement, $M$-theory,
on a $G_2$ manifold. Yet, until fairly recently, though $G_2$
manifolds attracted attention, some early references being
 \nref\gtwoa{S. Shatashvili and C. Vafa, ``Superstrings And
 Manifolds Of Exceptional Holonomy,'' hep-th/9407025.}%
\nref\gtwob{P. Townsend and G. Papadopoulos, ``Compactification
Of $D=11$ Supergravity On Spaces Of Exceptional Holonomy,''
hep-th/9506150, Phys. Lett. {\bf B357} (1995) 472.}%
\refs{\gtwoa , \gtwob},
 model-building based on them was
impractical. The reason is that if one assumes that $X$ is smooth
one runs into immediate difficulties, while if $X$ is not smooth,
supergravity is not valid and until recently one would not have
known how to proceed.

The difficulties for smooth $X$ are as follows. Since a compact
manifold of $G_2$ holonomy has no continuous symmetries,
continuous gauge symmetries in compactification on such a manifold
arise  purely from the three-form field $C$ of $M$-theory.  It
follows that the connected part of the gauge group is abelian, and
that all massless chiral superfields are neutral. These features
certainly clash with what we want for particle physics.

One way to get a non-abelian gauge group and chiral fermions from
$M$-theory is to compactify on a manifold with boundary
\ref\hw{P. Horava and E. Witten, ``Heterotic And Type I String
Dynamics From Eleven Dimensions,'' Nucl. Phys. {\bf B460} (1996)
506, hep-th/9510209. }. Here we want to go in a different
direction, using the fact that there are other ways to get gauge
fields and chiral fermions from singularities in geometry. For
example, gauge fields can arise from $A-D-E$ orbifold
singularities \ref\edwitten{E. Witten, ``Small Instantons In
String Theory,'' Mucl. Phys. {\bf B460} (1996) 541,
hep-th/9511030. }, and massless chiral superfields can arise from
conifold singularities \ref\astr{A. Strominger, ``Black Hole
Condensation And Duality In String Theory,'' Nucl. Phys. Proc.
Suppl. {\bf 46} (1996) 204.} or from $G_2$ analogs thereof that we
will recall later.

The starting point for getting nonabelian gauge symmetry in this
way is to consider $M$-theory on $\R^7\times \R^4/\Gamma$, where
$\Gamma$ is a finite group of symmetries of $\R^4$.  In fact, the
rotation group of $\R^4$ is $SO(4)=SU(2)_L\times SU(2)_R$, and we
take $\Gamma$ to be a subgroup of, say, $SU(2)_L$. Then $\Gamma$
acts freely on $\R^4$ except for a singularity at the origin
${\cal O}$, so $\R^7\times \R^4/\Gamma$ has the seven-dimensional
singular set $\R^7\times {\cal O}$.

The  claim now \edwitten\ is that in $M$-theory on this singular
space, nonabelian gauge fields propagate on $\R^7\times {\cal
O}$.  The gauge group, depending on the choice of $\Gamma$, can
be any of the groups $SU(N)$, $SO(2N)$, and $E_6$, $E_7, $ or
$E_8$. For example, to get gauge symmetry $SU(5)$, we take
$\Gamma$ to be $\Z_5$.

We have presented this so far as if the space on which the gauge
fields propagate is supposed to be $\R^7$, but in general that
might be compactified or partly compactified also.  What is
really important here is that gauge fields arise wherever there
is a singularity that looks locally like $\R^4/\Gamma$. For our
purposes, consider $M$-theory on $\R^4\times X$, where $X$ is a
singular manifold of $G_2$ holonomy.  Suppose that $X$ has a
singular set that looks locally like $Q\times \R^4/\Gamma$, for
some three-manifold $Q$. Then in $M$-theory compactification on
$\R^4\times X$, the singular set is $\R^4\times Q$.  If
$\Gamma=\Z_5$, we will get $SU(5)$ gauge fields on $\R^4\times Q$.

If moreover $Q$ has fundamental group $\Z_n$, then as in \chsw\
we can break $SU(5)$ to the standard model $SU(3)\times
SU(2)\times U(1)$ by taking the gauge field on $Q$ to be a flat
connection with a non-trivial holonomy.  In this way, we get a
model that has a standard model gauge group at low energies, with
low energy gauge couplings that obey the usual relation that
holds in $SU(5)$ supersymmetric GUT's, even though unification
only strictly holds in the higher dimension.

We still do not have chiral fermions.  In fact, if $Q$ is smooth
and $X$ has only the orbifold singularities that we have already
described, there will be no chiral fermions.  To get chiral
fermions in this construction, we should assume that $Q$ passes
through points at which the singularity of $X$ is ``worse'' than
an orbifold singularity.  Singularities of $X$ that will give a
$\5$ or $\bar\5$ of $SU(5)$ were described in \ref\aw{M. F.
Atiyah and E. Witten, ``$M$-Theory Dynamics On A Manifold Of
$G_2$ Holonomy,'' hep-th/0107177.}, and singularities that give a
$\10$ or $\bar \10$ were described in \baw. For our present
purposes, the details of these singularities are not important.
All that really matters is that in these constructions, one will
get chiral superfields, transforming in suitable representations
of $SU(5)$, and localized at points on $Q$.  Note that in these
constructions $Q$ itself is smooth but passes through a point of
$X$ at which $X$ has a ``bad'' singularity that generates charged
chiral supermultiplets.

Now we can conceive of many models, depending on the choice of
points on $Q$ at which chiral supermultiplets are localized, and
the quantum numbers of those multiplets. To solve the
doublet-triplet splitting problem, we must be somewhat more
specific.  We assume that $Q=\S^3/\Z_n$ and that $X$ has a global
$\Z_n$ symmetry $F$ which acts on $Q$ as described in \knonp. If
so, the singular points at which charged chiral superfields arise
will form $F$ orbits, and we will consider the case that these
points all lie in the fixed set $S_1\cup S_2$ of $Q$ that was
described in section 2.2.  Given this, we have all the ingredients
to construct models like those of section 2 in the $M$-theory
context: we asssume the presence of a ${ \5}$ of Higgs bosons on
$S_1$, a ${\bar\5}$ on $S_2$, and additional multiplets containing
standard model fermions on either $S_1$ or $S_2$ depending on
what we want to get.

As has been explained elsewhere \refs{\ewitten,\baw}, concrete
examples along these lines can be constructed using duality
between $M$-theory on K3 and the heterotic string on $\T^3$. (A
K3 surface is a four-manifold of $SU(2)$ holonomy.)  We start with
a heterotic string model involving compactification on a
six-manifold $Y$ of $SU(3)$ holonomy.  If $Y$ participates in
mirror symmetry (and it is believed that most manifolds of
$SU(3)$ holonomy do so) then in some limit of its moduli space,
$Y$ is fibered over some three-dimensional base $Q$ with generic
fiber $\T^3$ \ref\syz{A. Strominger, E. Zaslow, and S.-T. Yau,
``Mirror Symmetry Is $T$-Duality,'' Nucl. Phys. {\bf B479} (1996)
243, hep-th/9606040. }. By making fiber-wise the duality between
the heterotic string on $\T^3$ and $M$-theory on K3, we get an
equivalent $M$-theory model with compactification on a manifold
$X$ of $G_2$ holonomy that is fibered over $Q$ with generic fiber
K3.

If the heterotic string model has unbroken gauge symmetry $SU(5)$
(possibly broken globally by holonomies of a flat connection),
then the generic fiber of $X$ has a $\Z_5$ orbifold singularity
and the locus of these singularities is a copy of $Q$ on which
$SU(5)$ gauge fields propagate.    In this construction, if we
start in the heterotic string with a simply-connected
six-manifold $Y$, then $Q$ will be simply connected and will be,
in fact, a copy of $\S^3$.  But if $Y$ is not simply connected,
$Q$ will be a quotient of $\S^3$ such as   $Q=\S^3/\Z_n$, the
example considered in section 2.2. In this case, we can obtain in
the $M$-theory framework models similar to those of section 2.2,
solving the doublet-triplet splitting problem and achieving a
realistically long lifetime for the proton.

While clearly this sort of model is quite similar to the models
of section 2.2, there also are some differences. Some of these
were noted at the end of the introduction. We may here note some
additional differences:

(1)  In the context of deconstruction, we can build any model we
want; we have to impose anomaly cancellation by hand.  In the
$M$-theory framework, we are limited (in the choice of $n$, the
arrangment of singularities, etc.) by the difficult problem of
what singular manifolds of $G_2$ holonomy actually exist.  (For
an account of much of the present knowledge of construction of
smooth manifolds of $G_2$ holonomy, see \ref\joyce{D. Joyce, {\it
Compact Manifolds Of Special Holonomy} (Oxford Univ. Press,
2000). }.) This may mean that there is more predictive power in
principle, especially when one considers mechanisms of
supersymmetry breaking, but it is hard to extract it. It can be
shown that all singular $G_2$ manifolds are such that the gauge
anomalies cancel \ewitten.   In the deconstructed models based on
$SU(5)'\times SU(5)''$, one imposes anomaly cancellation as a
constraint, and one has to separately cancel gauge anomalies in
each factor.  In the $M$-theory models, there is an anomaly
inflow mechanism that can move anomalies from $S_1$ to $S_2$, and
triangle anomalies in general only  cancel when summed over all
multiplets supported on $S_1$ or $S_2$.

(2)  An $M$-theory model of the type we have described always has
an axion with a coupling to standard model gauge fields (and
conceivably solving the strong CP problem). This axion comes from
the mode $\int_QC$ of the $C$-field.  (Depending on the topology
of $X$, there may be other axion-like fields, not coupling to
standard model gauge fields but perhaps relevant to cosmology,
coming from other modes of $C$.  It is quite natural to have many
of these.) It is reasonable to expect by analogy with similar
results for continuous gauge symmetries \ewitten\ that in the
$M$-theory model, $F$ is always anomaly-free, if one allows in
general for a non-trivial transformation law of the axion under
$F$.  In the deconstructed models, there might not be an axion
and the global symmetry $F$ might be anomalous.

\nref\hm{J. A. Harvey and G. Moore,  ``Superpotentials And Membrane Instantons,''
JHEP {\bf 9809:004} (1998), hep-th/9808060.}%
(3) In the $M$-theory models, Yukawa couplings contributing to
quark and lepton masses always come from membrane instantons
\refs{\hm,\baw} (unless some of the points on $Q$ that support
chiral superfields coincide; then there may be classical
contributions).\foot{The membrane instanton generically intersects
$Q$ in a graph that connects special points at which chiral
superfields are supported. The amplitude for a membrane instanton
contribution to the superpotential ties together the various
superfields that enter the amplitude, and which may be supported
at different points on $Q$, by parallel transport along the
graph.  The symmetry-breaking holonomy $W$ may affect this
parallel transport, producing phases for the various instanton
contributions. } The quark and lepton masses can thus vary
exponentially with the instanton volume, and it is natural, if
some of the instantons have volume somewhat larger than the
$M$-theory scale, to get some extremely light quarks and
leptons.  To get a large mass, such as the top quark mass, we
have to assume that some singularities are nearby or perhaps even
coincident (giving classical contributions).   In the
deconstructed models, there is an analogous but somewhat
different mechanism, described in section 2.1, for suppressing
the masses of some quarks and leptons; this arises when some
fields can only receive mass as a result of unrenormalizable
couplings.

\bigskip\noindent{\it A Note On Yukawa Unification}

Finally, we note an interesting difference between this sort of
$M$-theory model and many other models considered in string
compactification.  GUT models sometimes lead to relations among
quark and lepton masses coming from unification \ref\nano{A. J.
Buras, J. Ellis, M. K. Gaillard, and D. Nanopoulos, ``Aspects Of
The Grand Unification Of Strong, Weak, and Electromagnetic
Interactions,'' Nucl. Phys. {\bf B159} (1979) 16.}. For example,
if the $b$ quark and $\tau$ lepton arise from the same $\bar \5$
and $\10$ of $SU(5)$, then the Yukawa couplings giving rise to
their masses are related by $SU(5)$ symmetry, giving a relation
between them that is often called Yukawa unification. This leads
to a relation between the $b$ and $\tau$ masses which, after
allowing for renormalization from the GUT scale to the weak
scale, agrees pretty well with experiment.

The analogous relations for the first two generations do not work
well, however, and before going on let us note a simple way to
avoid this problem in four-dimensional $SU(5)$ models.  We simply
add an extra $\5'\oplus \bar \5'$ of $SU(5)$, with GUT scale
masses, but mixing with the $\bar \5\oplus \10$ of, say, the first
generation, in such a way that the observed down quark and
electron are not $SU(5)$ partners.  To explain this in detail, we
denote by $\Phi$ a Higgs multiplet in the adjoint representation
of $SU(5)$ whose expectation value $\langle \Phi\rangle$ breaks
$SU(5)$ to the standard model.  We derive  the mixing structure
from generic renormalizable couplings \eqn\polo{M\bar
\5'\5+\Phi(a\bar \5'+b\bar \5)\5} with mass $M$ and couplings
$a,b$; we suppose $M\sim a\Phi\sim b\Phi$. Given such a
structure, out of the $\bar \5$ and $\bar \5'$, one field with
down quark quantum numbers and one with electron quantum numbers
pair up with the $\5$ to get a GUT mass. But because of the
coupling to $\Phi$, the heavy components of the $\bar \5$ and
$\bar \5'$ are not $SU(5)$ partners -- different linear
combinations are massive for down quarks and leptons. Likewise,
one field with down quark quantum numbers and one field with
electron quantum numbers will escape having GUT masses, but they
will also not be $SU(5)$ partners. Hence their Yukawa couplings
to Higgs bosons are not related by $SU(5)$ and the usual Yukawa
unification will not hold. (Likewise many standard estimates
concerning proton decay will be modified by this mixing.)  Since
standard Yukawa unification does work for the third generation,
one might want to impose discrete symmetries such that (to some
approximation) this sort of mixing only affects the first two
generations.

In many string theory compactifications, this sort of mixing
occurs (with the role of the heavy fields being played by
Kaluza-Klein excitations or massive string modes) and the
standard Yukawa unification does not hold. For example, in the
original Calabi-Yau model \chsw, despite its to similarity to
four-dimensional GUT's, the usual GUT relations among Yukawa
relations are not valid: the unified group is only present in ten
dimensions, and by the time one reduces to four dimensions, one
cannot make sense of the question of which down quark is the
$SU(5)$ or $E_6$ partner of the electron. By contrast, in the
$M$-theory models discussed here, even though $SU(5)$ is broken,
the special points in $Q$ support complete $SU(5)$ multiplets. So
there is always a natural notion of which down quark is the
$SU(5)$ partner of a given charged lepton, namely the one that is
supported at the same point on $Q$.

Whether the Yukawa couplings obey $SU(5)$ relations is a more
subtle question.  Since the Yukawa couplings come from nonlocal
effects (membrane instantons), they might conceivably see $SU(5)$
breaking by Wilson lines.  However, to the extent that each
Yukawa coupling comes mainly from a single type of instanton, the
Wilson lines, which contribute phases to the amplitudes (by a
mechanism mentioned in the last footnote), would not disturb the
$SU(5)$ mass relations. (Wilson lines might give different phases
to different membrane instantons contributing to a given Yukawa
coupling, and thereby spoil $SU(5)$ relations among the couplings
if more than one instanton contributes.)   So in this situation
we do have a framework for why some quark and lepton masses might
obey approximate $SU(5)$ relations.

\bigskip
This work was supported in part by NSF Grant PHY-0070928.
\listrefs
\end